\documentclass[11pt]{cernrep}
\usepackage{amssymb,amscd}
\usepackage[figuresright]{rotating}
\usepackage{epsfig}
\usepackage{array}
\usepackage{graphics}
\usepackage{graphicx}
\usepackage{multirow}
\usepackage{supertabular}
\usepackage{longtable}
\usepackage{dcolumn}
\usepackage{xspace}
\usepackage{tabularx}
\usepackage{color}
\usepackage{calc}
\usepackage{floatflt}
\usepackage{cite,./mcite}

\begin{document}
\title{The structure of the proton and NLO QCD fits}                            
\author{Claire Gwenlan, on behalf of the H1 and ZEUS Collaborations }
\institute{Nuclear \& Astrophysics Laboratory, Keble Road, Oxford, OX1 3RH. UK.}
\maketitle

\begin{abstract}
HERA has provided a wealth of high precision structure function and jet production data, allowing considerable progress to be made in understanding the structure of the proton. In this paper, several of the most recent proton structure results from the H1 and ZEUS Collaborations at HERA are presented. These include results from NLO QCD fits, neutral current deep inelastic scattering and jet production at high tranvsverse energy.
\end{abstract}

\section{Introduction}
This paper reports on several of the most recent results on proton structure from HERA. The ZEUS Collaboration have performed a new next-to-leading order (NLO) QCD fit to both structure function and jet data to determine the proton parton distribution functions (PDFs) and the strong coupling, $\alpha_s$. ZEUS have also performed a 
measurement of neutral current deep inelastic scattering (DIS), introducing a new method which allows differential cross section measurements up to Bjorken $x$ of 1. Finally, the H1 Collaboration have performed new measurements of dijets in photoproduction, which may provide additional constraints on the gluon distribution in the proton.  

\section{HERA physics and kinematics}                                          
The kinematics of lepton-proton DIS are described 
in terms of the Bjorken scaling variable, $x$, the negative invariant 
mass squared of the exchanged vector boson, $Q^2$, and the 
fraction of energy transferred from the lepton to the hadron system, $y$. 
At leading order (LO) in the electroweak interaction, the double 
differential cross section for the neutral (NC) and charged (CC) current processes 
are given in terms of proton structure functions,
$\frac{{\rm d}^2\sigma(e^{\pm}p)}{{\rm d}x{\rm d}Q^2}= \left [
Y_+ {\rm F}_2 - y^2 {\rm F_L} \mp Y_- x{\rm F}_3 \right ],$  
where $Y_{\pm}=1\pm(1-y)^2$. 
The structure functions are directly related to the proton PDFs and 
their $Q^2$ dependence, or scaling violation, is predicted in perturbative QCD. 
The QCD scaling violations in the inclusive cross section data, 
namely the QCD Compton ($\gamma^{*}q \rightarrow gq$) and 
boson-gluon-fusion ($\gamma^{*}g \rightarrow q\bar{q}$) 
processes, may also give rise to distinct jets in the final state.
Jet cross sections therefore provide a direct 
constraint on the gluon through the boson-gluon-fusion process.                

\section{Recent results on proton structure from HERA}
\subsection{NLO QCD fits using HERA data}
The ZEUS Collaboration have recently performed a new NLO QCD fit~\cite{epj:c42:1}, to their full set of HERA-I (94-00) NC and CC inclusive data~\cite{epj:c21:443,epj:c28:175,hep-ex:0401003,epj:c12:411,pl:b539:197,epj:c32:16}, as well as to high precision jet data in inclusive jet DIS~\cite{pl:b547:164} and dijet photoproduction~\cite{epj:c23:615}. 
This is called the ZEUS-JETS fit. 

The low $Q^2$ NC data determine the low-$x$ sea and gluon distributions, while the high-$Q^2$ NC and CC data constrain the valence quarks. The jet data directly constrain the gluon in the mid-to-high-$x$ region ($x \approx 0.01-0.5$). This is the first time that jet data have been {\it rigorously} included in a QCD fit.
The use of only ZEUS data eliminates the uncertainties from heavy-target corrections 
that are present in global analyses, which also include fixed target data. It also avoids the complications that can sometimes arise 
from combining data-sets from different experiments, thereby allowing a rigorous statistical treatment of the experimental uncertainties. Full details of the ZEUS-JETS PDF parameterisation and assumptions of the fit are given elsewhere~\cite{epj:c42:1}.

\begin{figure}[Ht]
{\hspace{-0.4cm}
{\includegraphics[width=8.5cm,height=8.5cm]{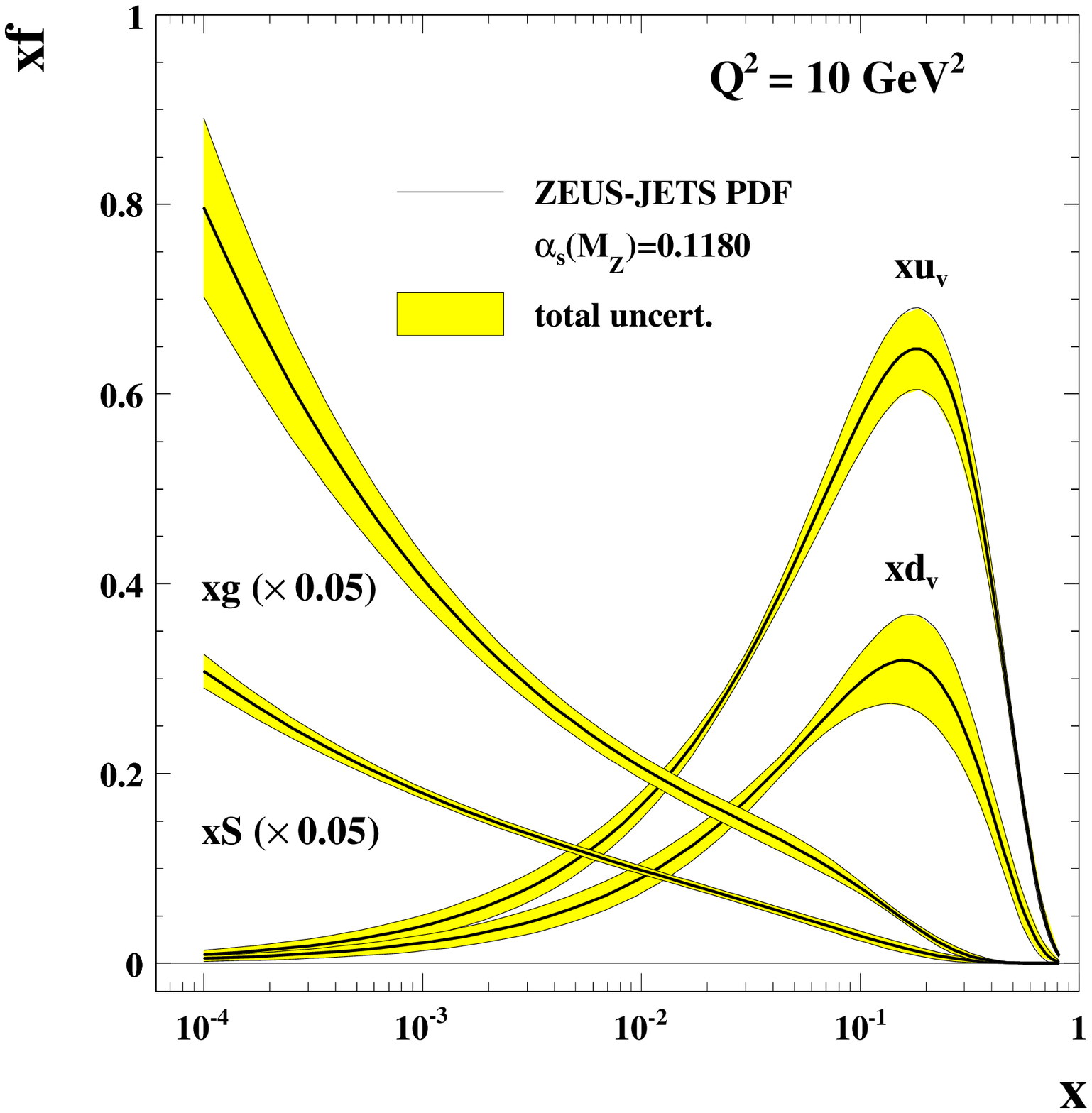}\hspace{-0.4cm}\includegraphics[width=7.5cm,height=8.2cm]{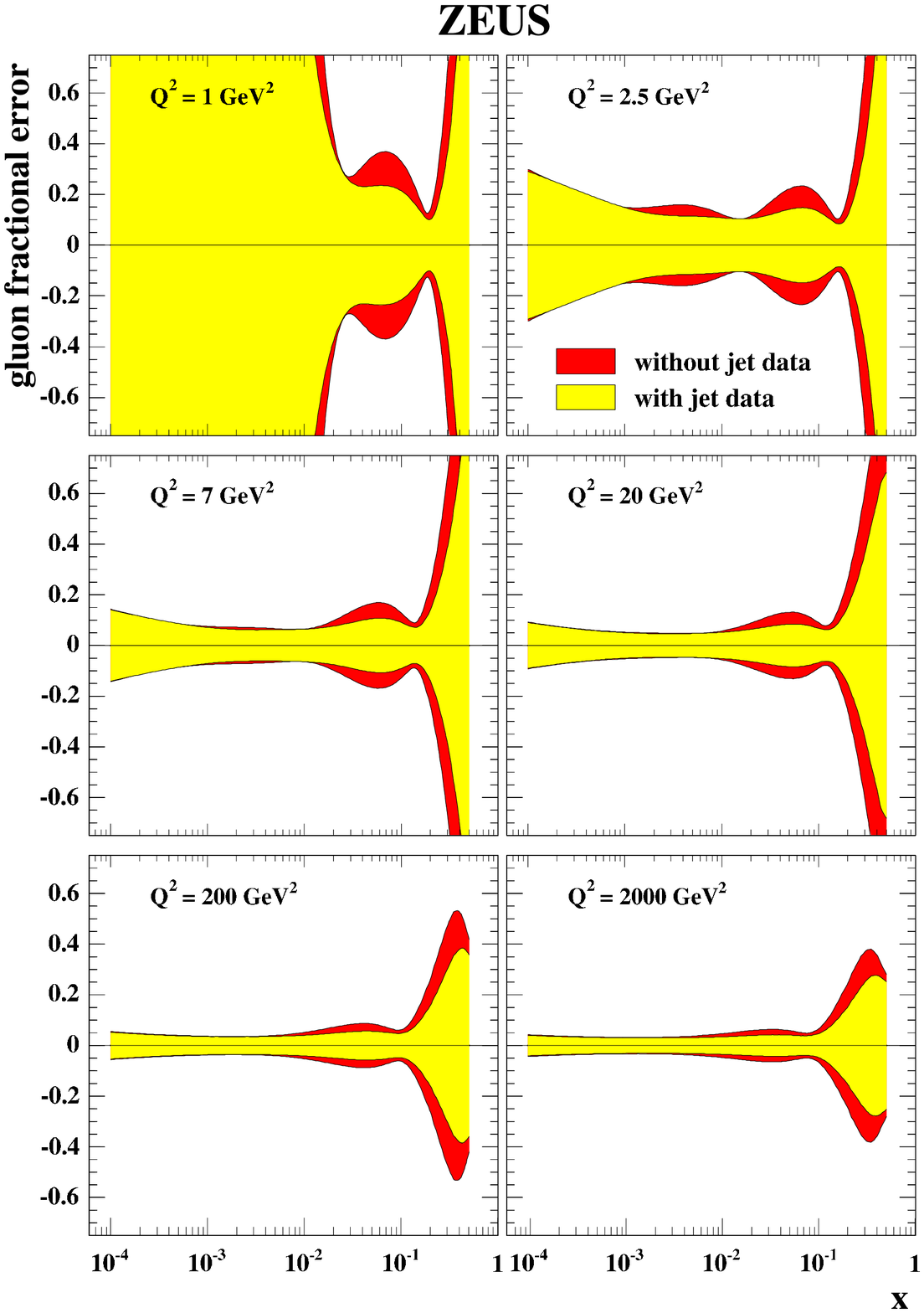}}}
\caption{Left: $u$-valence, $d$-valence, sea and gluon PDFs from the ZEUS-JETS fit, for $Q^2=10$ ${\rm GeV}^2$. Right: fractional uncertainty on the gluon PDF for different $Q^2$ values, for fits with (yellow) and without (red) jet data included.}
\label{Fig:PDF}
\end{figure}

The $u$-valence, $d$-valence, sea and gluon PDFs extracted from the ZEUS-JETS fit are shown for $Q^2=10$ ${\rm GeV}^2$ in Fig.~\ref{Fig:PDF} (left). 
The PDFs produce a good description of both the inclusive and jet cross section data from HERA, demonstrating the validity of QCD factorisation.
The jet data are directly sensitive to the gluon distribution through the boson-gluon-fusion process. Figure~\ref{Fig:PDF} (right) shows the fractional uncertainty on the gluon density for a range of $Q^2$ values, for the ZEUS-JETS fit (yellow) and a similar fit without jet data (red). The inclusion of jet data is shown to significantly improve the gluon uncertainties at mid-to-high-$x$. This improvement persists to high scales.\vspace{0.2cm} \\ 
{\bf Extraction of $\alpha_s(M_Z)$}\\
In the inclusive cross sections, $\alpha_s$ and the gluon are strongly correlated. Jet production through the boson-gluon-fusion process directly depends on the gluon PDF. However, the QCD Compton process depends only on the quark distribution. Hence, the use of jet data in the fit allows a precise extraction of $\alpha_s$, without a strong correlation to the gluon.
The value of $\alpha_s(M_Z)$ has been determined from the ZEUS-JETS fit by treating it as an additional free parameter. The value extracted is:
$\alpha_s(M_Z) = 0.1183 \pm 0.0028 ~{\rm (exp.)}  
                 \pm 0.0008 ~{\rm (model)} \pm 0.005 ~{\rm (scale)}$. This 
is in good agreement with the world average of $0.1182 \pm 0.0027$.

\subsection{Neutral current cross sections at high-$x$ at HERA}
HERA has provided a wealth of precision structure function data at low Bjorken $x$. However, the high-$x$ region 
remains largely unexplored, due to limitations in beam energies and measurement techniques. 
The ZEUS Collaboration have developed a new method to select and measure events in NC DIS at very high $x$, 
at intermediate $Q^2$ ($Q^2 > 576$ ${\rm GeV}^2$). These events are characterised by a well reconstructed, 
high energy electron (or positron)\footnote{The term ``electron'' will be used from now on to denote either an 
electron or a positron, unless specifically stated.} in the central part of the calorimeter, and a jet from the 
struck quark. As $x$ increases, the jet is boosted more and more forwards and is eventually lost down the beam-pipe. 
The value of $x$ at which this occurs is $Q^2$ dependent.
\begin{figure}[Ht]
\includegraphics[width=15.cm,height=8.5cm]{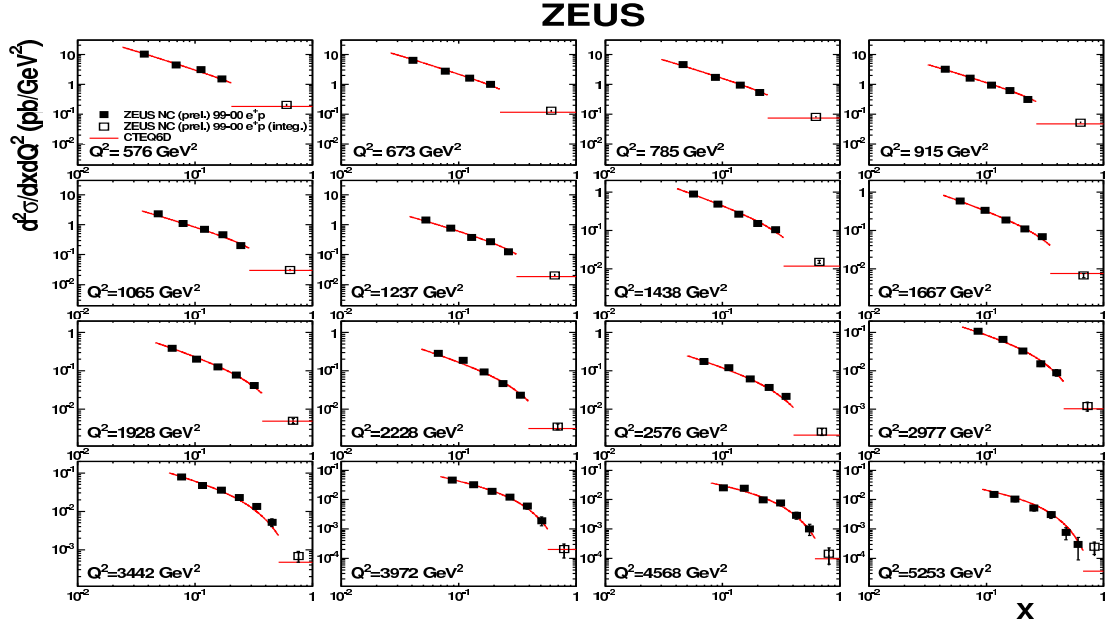}
\caption{Double differential cross sections (solid squares) for the 99-00 $e^+p$ data, as a function of $x$ in $Q^2$ bins, compared to the CTEQ6D proton PDFs. The last bin (open symbol) shows the integrated cross section over $x$, divided by the bin width $1/(1-x_{\rm edge}).\int_{x_{\rm edge}}^1 ({\rm d}^2\sigma/{\rm d}x{\rm d}Q^2){\rm d}x$; the symbol is shown at the centre of the bin. In this bin the predictions are shown as a horizontal line. The error bars show the quadratic sum of the statistical and systematic uncertainties.}
\label{Fig:NC}
\end{figure}

The new method combines electron and jet information to give the best possible reconstruction of the event 
kinematics, as follows. First, the $Q^2$ of the event is reconstructed using information from the scattered lepton 
: $Q^2 = 2E_e E_e' (1+\cos\theta_e)$, where $E_e$ is the lepton energy and $\theta_e$ is the lepton scattering angle. 
Events are then separated into those with exactly one well reconstructed jet (satisfying $E_{\rm T}^{jet} > 10$ GeV and 
$\theta^{jet}>0.12$) and those with zero jets. For the 1-jet events, the jet information is used to calculate the value 
of $x$ from $E^{jet}$ and $\theta^{jet}$. This allows a measurement of the double differential cross section 
${\rm d}^2\sigma/{\rm d}x{\rm d}Q^2$. For the 0-jet sample, events are assumed to come from high $x$, and are 
collected in a bin with $x_{\rm edge} < x < 1$. The value of $x_{\rm edge}$ is evaluated for each $Q^2$ bin, based on kinematic constraints. An integrated cross section is then calculated according to: $\int_{x_{\rm edge}}^1 
({\rm d}^2\sigma/{\rm d}x{\rm d}Q^2) {\rm d}x$. The features of this method are: i) good resolution in  $Q^2$ for all $x$, ii) good resolution in $x$ for events where a jet can be reconstructed and iii) differential cross section measurements possible up to $x=1$.
  
The NC cross sections have been measured using the new method for 16.7 ${\rm pb}^{-1}$ 98-99 $e^-p$ and 65.1 ${\rm pb}^{-1}$ 99-00 $e^+p$ data. The results for the 99-00 data-set are 
shown in Fig.~\ref{Fig:NC}, compared to the Standard Model expectations at NLO, using the CTEQ6D~\cite{cteq6d} proton PDFs. 
The double differential cross sections are shown by the solid points. The data are generally well described by the 
Standard Model predictions. The open squares show the integrated cross sections, in bins ranging from $x_{\rm edge}$ to $1$. 
The precision of the integrated cross section points are comparable to the other points. For most of the highest-$x$ bins, where no previous measurements exist, the data tend to lie above the expectation from CTEQ6D. These 
measurements will provide new constraints on the 
valence quark PDFs at high $x$.

\subsection{Dijets in photoproduction at high-$E_{T}$ at HERA}
The H1 Collaboration have performed a new measurement of dijets in photoproduction; a process in which the beam electron interacts with the proton via the exchange of an almost real photon ($Q^2 \approx 0$). Photoproduction can be separated into two contributions: {\it direct processes}, in which the photon itself participates in the hard scatter, contributing a fraction $x_\gamma =1$ of the photon's longitudinal momentum, and {\it resolved processes}, in which the photon first fluctuates into partons, each carrying a fraction $x_\gamma < 1$ of the photon's momentum, and one of these participates in the hard interaction. In the latter case, the hadronic structure of the photon is described by associated parton density functions. 
The results presented here represent an update to a previous publication~\cite{epj:c25:13}, with twice the statistics and an improved understanding of the systematic uncertainties on the measurement. In addition, new cross sections with different jet topologies have also been measured. The goal is to provide the best possible measurement in order to provide additional constraints on the gluon PDF in the proton and $\alpha_s$ in future QCD fits to H1 data.
\begin{figure}[Ht]
\includegraphics[width=15cm,height=9.7cm]{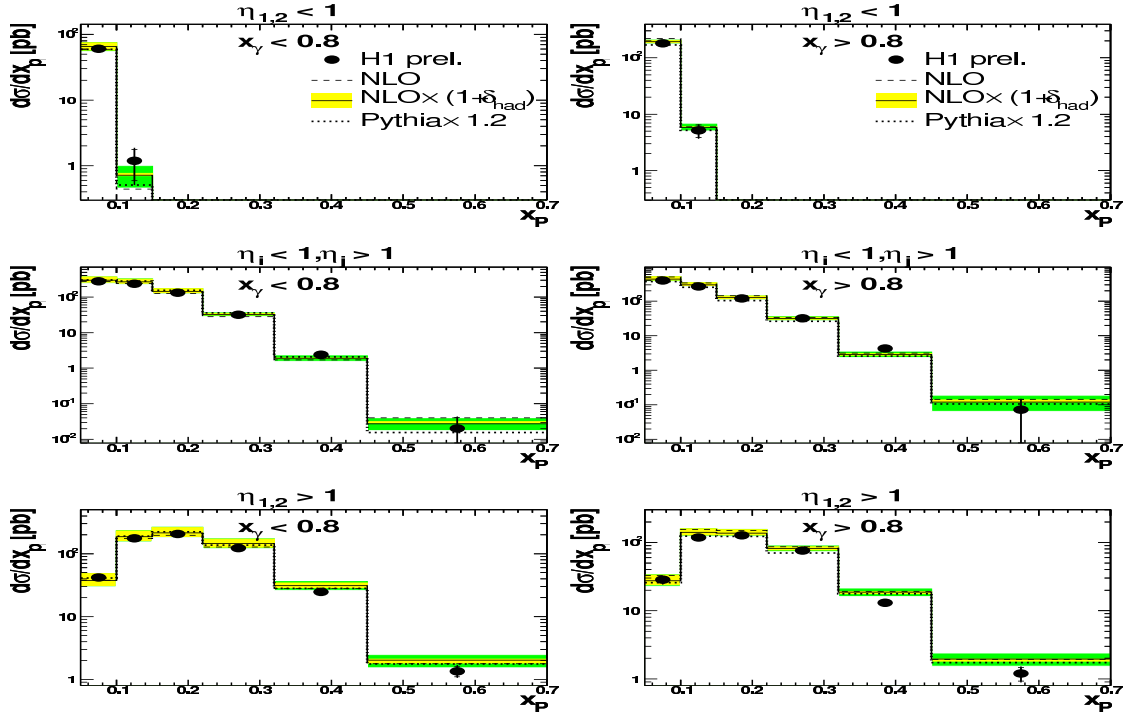}
\caption{Cross sections differential in $x_p$ with different topologies of jet $\eta$ for data (points), NLO QCD with (solid line) and without (dashed) hadronisation corrections $\delta_{had}$ and for PYTHIA (dotted) scaled by a factor of 1.2. The inner error bars on the data points show the statistical uncertainty and the outer error bars show the total experimental uncertainty. The inner band of the NLO$\times(1+\delta_{had})$ prediction shows the scale uncertainty and the outer band is the total theoretical uncertainty including contributions from the PDFs. The cross sections are shown for two regions in $x_{\gamma}$, enhancing the resolved (left) and direct (right) contributions.}
\label{Fig:Dijet}
\end{figure}

The dijet cross sections have been measured using 66.6 ${\rm pb}^{-1}$ $e^+p$ data taken in 99-00, in the kinematic region defined by: $Q^2 < 1$ ${\rm GeV}^2$, $0.1<y<0.9$, $p_{\rm T}^{jet1,2} > 25,15$ GeV and $-0.5 < \eta^{jet} < 2.75$. In Fig.~\ref{Fig:Dijet}, the cross sections are shown as a function of the proton fractional momentum, $x_p$, separated into two regions of $x_{\gamma}$ and in different regions of jet $\eta$. The low-$x_\gamma$ region corresponds to events which are resolved-enriched, while the high-$x_\gamma$ region corresponds to those which are direct-enriched. The data are compared to the predictions of NLO QCD~\cite{fr} and to the PYTHIA~\cite{pythia} Monte Carlo. The data are generally well described by the predictions although there is a tendency for the NLO prediction to overshoot the data when both jets are in the forward direction. The uncertainties from PDFs are smaller at low $x_p$, since in this region they are already well constrained from inclusive DIS data. However, the experimental and scale uncertainties are larger at low  $x_p$ than at high $x_p$. In the region $x_\gamma > 0.8$, where the photon interacts directly with the proton, there is little dependence on photon structure. Therefore, this is an ideal facility to test the structure of the proton. The experimental and scale uncertainties are also smaller in this region. 
These data should allow studies of the impact of the jet data on PDFs in future combined QCD fits of these data with inclusive measurements.

\end{document}